\documentclass[aps,prb,superscriptaddress,floatfix,preprint]{revtex4-1}

\usepackage{amsmath}
\usepackage{amsfonts}
\usepackage{amssymb}
\usepackage{epsfig}
\usepackage{amssymb}
\usepackage{graphicx}
\usepackage{color}
\usepackage{siunitx}

\begin{document}

\title{Gapless Andreev bound states in the quantum spin Hall insulator HgTe}

%

\author{E.~Bocquillon}
\thanks{All three authors contributed equally to this work, email: erwann.bocquillon@physik.uni-wuerzburg.de}
\affiliation{Physikalisches Institut (EP3), Universit\"at W\"urzburg, Am Hubland, D-97074 W\"urzburg, Germany}

\author{R.S.~Deacon}
\thanks{All three authors contributed equally to this work, email: erwann.bocquillon@physik.uni-wuerzburg.de}
\affiliation{Advanced Device Laboratory, RIKEN, 2-1 Hirosawa, Wako-shi, Saitama, 351-0198, Japan}
\affiliation{Center for Emergent Matter Science, RIKEN, 2-1 Hirosawa, Wako-shi, Saitama, 351-0198, Japan}

\author{J.~Wiedenmann}
\thanks{All three authors contributed equally to this work, email: erwann.bocquillon@physik.uni-wuerzburg.de}
\affiliation{Physikalisches Institut (EP3), Universit\"at W\"urzburg, Am Hubland, D-97074 W\"urzburg, Germany}

\author{P.~Leubner}
\affiliation{Physikalisches Institut (EP3), Universit\"at W\"urzburg, Am Hubland, D-97074 W\"urzburg, Germany}

\author{T.M.~Klapwijk}
\affiliation{Kavli Institute of Nanoscience, Faculty of Applied Sciences, Delft University of Technology, Lorentzweg 1, 2628 CJ Delft, The Netherlands}

\author{C.~Br\"une}
\affiliation{Physikalisches Institut (EP3), Universit\"at W\"urzburg, Am Hubland, D-97074 W\"urzburg, Germany}

\author{K.~Ishibashi}
\affiliation{Advanced Device Laboratory, RIKEN, 2-1 Hirosawa, Wako-shi, Saitama, 351-0198, Japan}
\affiliation{Center for Emergent Matter Science, RIKEN, 2-1 Hirosawa, Wako-shi, Saitama, 351-0198, Japan}

\author{H.~Buhmann}
\affiliation{Physikalisches Institut (EP3), Universit\"at W\"urzburg, Am Hubland, D-97074 W\"urzburg, Germany}

\author{L.W.~Molenkamp}
\affiliation{Physikalisches Institut (EP3), Universit\"at W\"urzburg, Am Hubland, D-97074 W\"urzburg, Germany}

\maketitle

\noindent

\newpage


{\bf In recent years, Majorana physics has attracted considerable attention in both theoretical and experimental studies due to exotic new phenomena and its prospects for fault-tolerant topological quantum computation. To this end, one needs to engineer the interplay between superconductivity and electronic properties in a topological insulator, but experimental work remains scarce and ambiguous. Here we report experimental evidence for topological superconductivity induced in a HgTe quantum well, a two-dimensional topological insulator that exhibits the quantum spin Hall effect. The ac Josephson effect demonstrates that the supercurrent has a $4\pi$-periodicity with the superconducting phase difference as indicated by a doubling of the voltage step for multiple Shapiro steps. In addition, an anomalous SQUID-like response to a perpendicular magnetic field shows that this $4\pi$-periodic supercurrent originates from states located on the edges of the junction. Both features appear strongest when the sample is gated towards the quantum spin Hall regime, thus providing evidence for induced topological superconductivity in the quantum spin Hall edge states.}

\begin{figure}[h!]
\centerline{\includegraphics[width=0.9\textwidth]{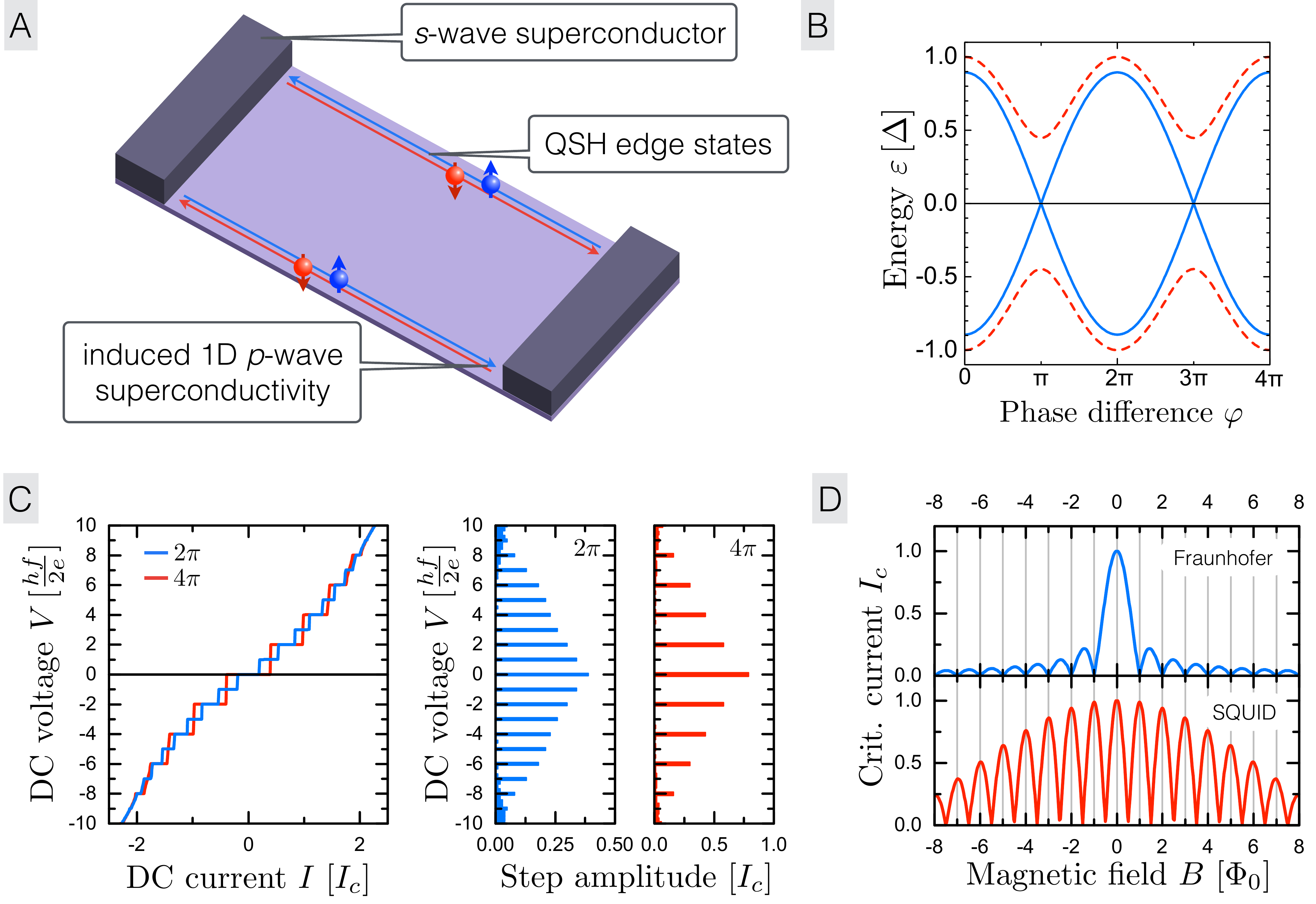}}
\caption{{\bf Physics in a topological Josephson junction -} A) Schematics of a topological Josephson junction with induced $p$-wave superconductivity. A superconducting weak link in the QSH regime contacted by two superconducting $s$-wave electrodes hosts induced $p$-wave superconductivity. B) Andreev spectrum of a topological $p$-wave Josephson junction (in the short junction limit). The Andreev bound states, located on the edges of the samples, have a protected crossing at zero energy and a $4\pi$-periodicity in the superconducting phase difference $\varphi$ (blue line), in contrast with conventional $2\pi$-periodic Andreev bound states (red dashed line). C) Simulated $I$-$V$ curves in the presence of rf excitation for a $2\pi$-periodic (blue line) and $4\pi$-periodic (red line) supercurrents. Shapiro steps of quantized voltages $V_n=nhf/2e$ (with $n$ integer) occur in the first case, but only the sequence of even steps is visible in the latter case. Histograms of the voltage distribution (in bins of $0.25hf/2e$) indicates the presence of a given Shapiro step as a peak in the histogram. The two bar plots for the two previous curves highlight the absence of odd steps in the case of $4\pi$-periodicity. D) Simulated critical current $I_c$ as a function of the magnetic field (in units of the number of flux quanta through the junction area). For a uniform planar current, a Fraunhofer pattern (blue line) is depicted. For current flowing on the edges, a (dc) SQUID pattern (red line) is expected.} \label{Fig:1TheoryDevice}
\end{figure}

The realization of Majorana bound states is theoretically expected in a one-dimensional $p$-wave superconducting phase without spin-degeneracy\cite{Kitaev2001,Kwon2004}. A convenient experimental implementation of this exotic phase can be obtained by combining recently discovered topological states with conventional $s$-wave superconductivity\cite{Alicea2012,Beenakker2013}. Most of the experimental focus has been to date on 1D InAs or InSb nanowires which may undergo a topological phase transition under an appropriate applied magnetic field.  Although first results\cite{Mourik2012, Rokhinson2012} have been obtained, the topological origin of the observed phenomena remains unclear partly because the helical transport in the normal state has not been demonstrated.  A potential alternative platform is provided by quantum spin Hall insulators, in which electrons flow in two counter-propagating one-dimensional edge states of opposite spins\cite{Bernevig2006,Konig2007} (see Fig.\ref{Fig:1TheoryDevice}A). Unlike nanowires, this topological state is present in the absence of magnetic field, thus alleviating the requirements for high critical field superconductors\cite{Samkharadze2015}. Ideally, a Josephson junction formed from a quantum spin Hall (QSH) insulator and conventional $s$-wave superconducting contacts is expected to emulate spinless $p_x+ip_y$-wave superconductivity\cite{Fu2009}. On each edge, it contains one Andreev doublet with a topologically protected crossing for a superconducting phase difference $\varphi=\pi$ (Fig.\ref{Fig:1TheoryDevice}B). The two states of this gapless topological Andreev doublet (usually called Majorana bound states) have a $4\pi$-periodicity in the superconducting phase difference $\varphi$ and can thus carry a $4\pi$-periodic supercurrent $I_{4\pi}\sin\varphi/2$ along the edges of the sample\cite{Beenakker2013a}. This contrasts with conventional $2\pi$-periodic Andreev bound states, carrying a current $I_{2\pi}\sin\varphi$ (+higher harmonics). This theoretical expectation for unconventional Josephson effect motivates our experiment.

\begin{figure}[h!]
\centerline{\includegraphics[width=1\textwidth]{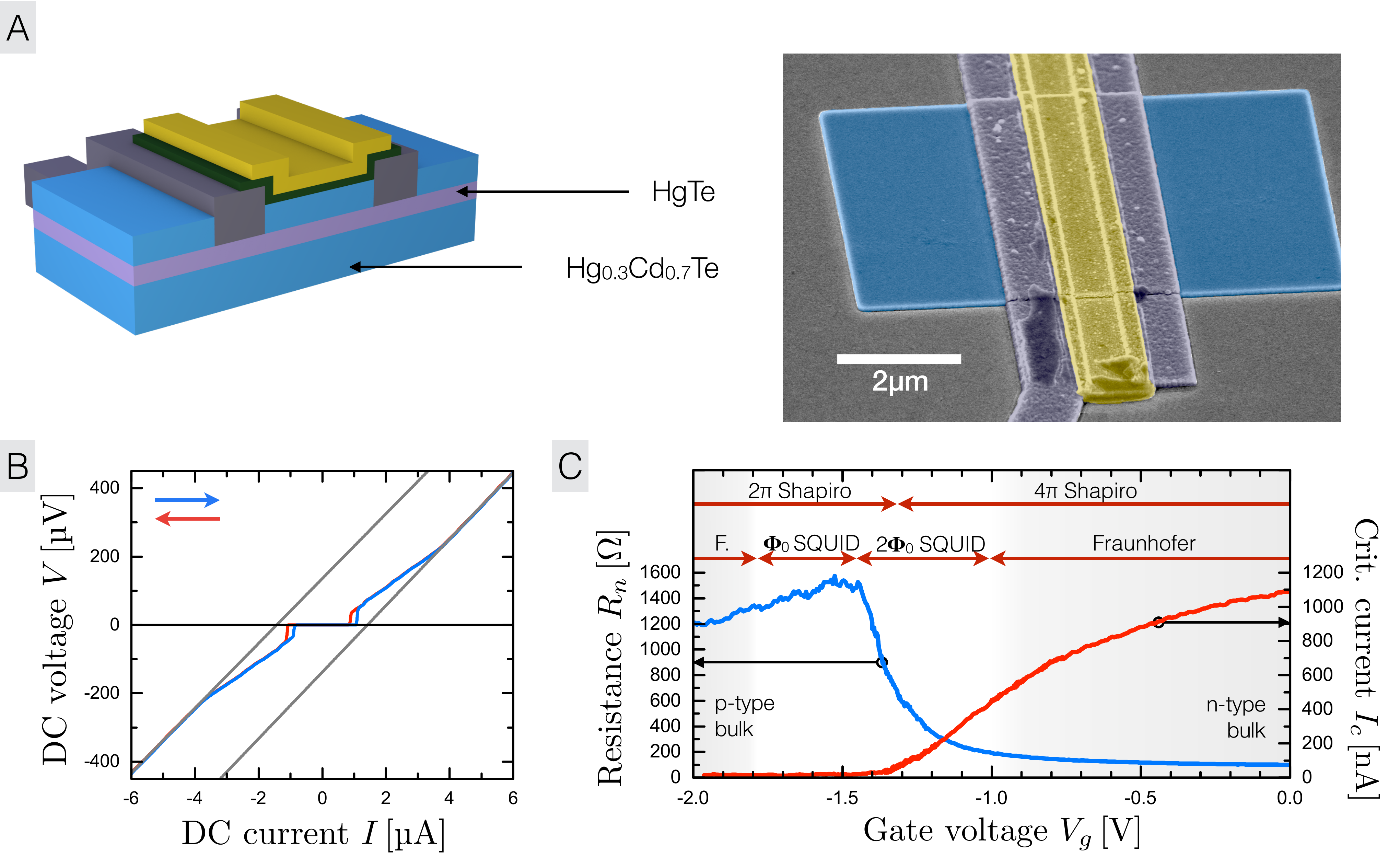}}
\caption{{\bf Experimental realization of a topological Josephson junction -} A) Artist view and colorized SEM picture of a junction. The HgTe QSH insulator (in mauve) is sandwiched between two layers of Hg$_{0.3}$Cd$_{0.7}$Te (in blue). The Al superconducting contacts are in dark purple while the gate is in yellow and lies on a thin dielectric layer of HfO$_2$ (dark green). B) $I$-$V$ curve measured at a gate voltage of $V_g=\SI{0}{\volt}$. It exhibits a critical current $I_c=\SI{1.1}{\micro\ampere}$, with a weak hysteresis visible between forward and reverse sweep (blue and red lines). For high biases, the asymptotes (grey lines) yield the normal state $R_n$ and signals the presence of an excess current $I_{exc}$. D) Critical current $I_c$ (red line), and normal state resistance $R_n$ (blue line) as a function of gate voltage $V_g$. The red arrows summarize the ranges where we observe the anomalous Josephson effect properties.} \label{Fig:2ExpDevice}
\end{figure}

\begin{figure}[h!]
\centerline{\includegraphics[width=1\textwidth]{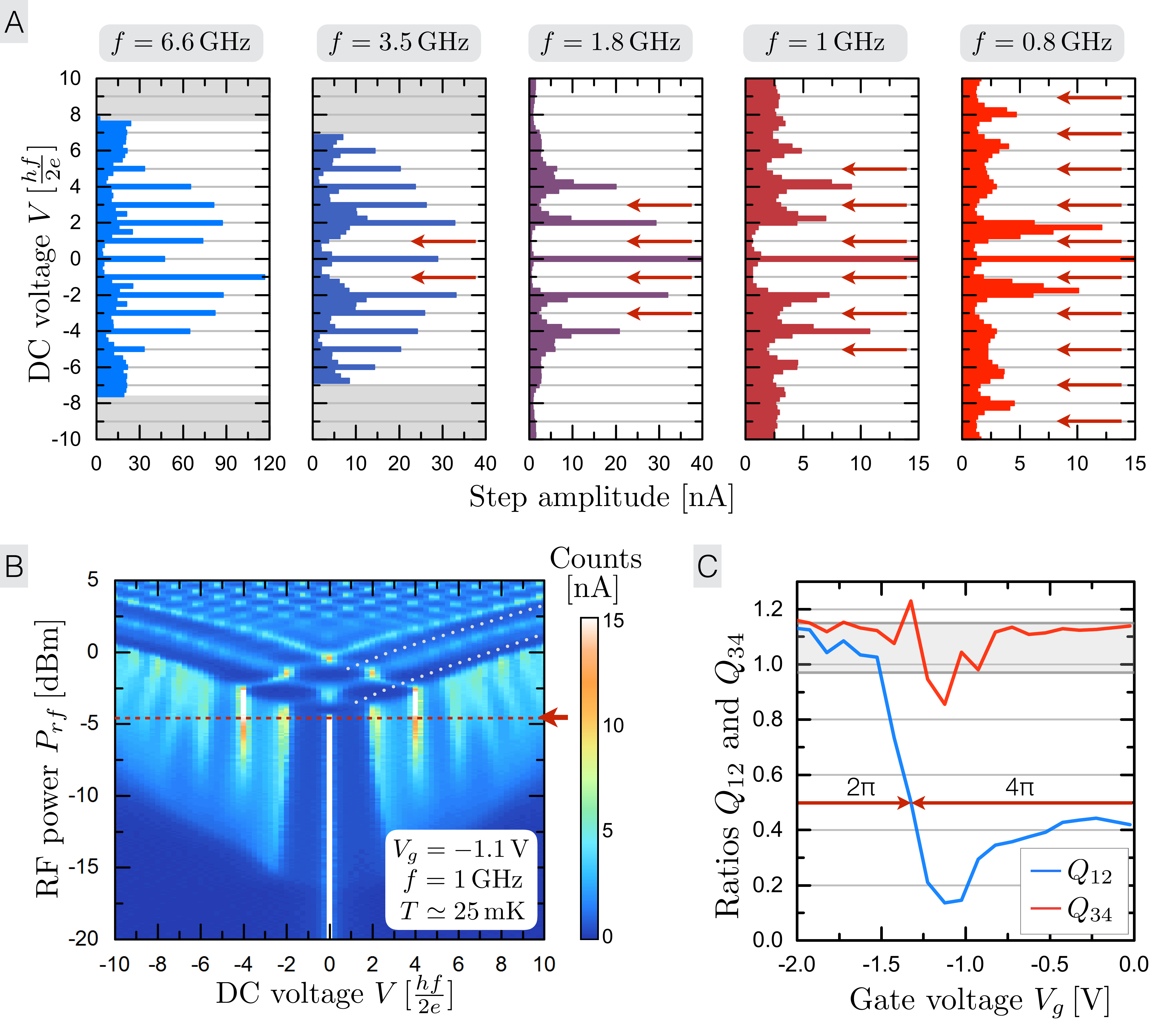}}
\caption{{\bf Response to an rf excitation -} 
A) Histograms of the voltage distribution obtained for different frequencies. For a high frequency $f=\SI{6.6}{\giga\hertz}$, all steps are present. For lower frequencies, we observe the disappearance of steps $n=1$ and 3 ($f=\SI{1.8}{\giga\hertz}$), and up to $n=9$ at $f=\SI{0.8}{\giga\hertz}$. Missing steps are highlighted by red arrows. B) Map of the voltage distributions with voltage bins in normalized units $hf/2e$ indicated on the abscissa and the RF excitation power on the ordinate. Steps $n=1,3$ are completely suppressed in the low power regime. In the oscillatory high power regime, dark fringes (white dotted lines) develop as the first and third oscillations are suppressed. The red dashed line indicates the linecut taken in the low power regime used to compute the histogram in A). C) Ratios of step amplitudes $Q_{12}$ (blue line) and $Q_{34}$ (red line), measured for $f=\SI{3.1}{\giga\hertz}$ as a function of gate voltage $V_g$. Both ratios exhibit a dip in the range \SI{-1.3}{\volt}-\SI{-0.9}{\volt}.} \label{Fig:3Shapiro}
\end{figure}

Here we report on the realization of a device that follows the proposal of Fu and Kane\cite{Fu2009} using HgTe, the first material to be identified as a topological insulator\cite{Konig2007}. Due to their inverted band structure\cite{Bernevig2006}, HgTe quantum wells of suitable thickness are a quantum spin Hall insulator in which superconductivity can be induced by means of e.g. Al electrodes\cite{Hart2014}. The anticipated presence of gapless Andreev bound states on the edges of such a device should be evidenced by two remarkable signatures, which we present as simulations in Fig.\ref{Fig:1TheoryDevice} (C and D). First, a $4\pi$-periodic supercurrent is expected in the ac Josephson effect (Fig.\ref{Fig:1TheoryDevice}C). When phase locking occurs between the junction dynamics and an external rf excitation, Shapiro steps\cite{Shapiro1963} appear at discrete voltages given by $V=nhf/2e$, where $n$ is the step index (blue line, left panel). In the presence of a $4\pi$-periodic supercurrent, an unconventional sequence of even steps (red line), with missing odd steps, is expected, reflecting the doubled periodicity of the Andreev bound states\cite{SanJose2012,Houzet2013,Badiane2013}. The exact sequence of visible steps can be highlighted by plotting a histogram of the voltage distribution as presented (right panels). In this research, we report on the experimental observation of an even sequence, with missing odd steps up to $n=9$ (Fig.\ref{Fig:3Shapiro}). The estimated amplitude of the $4\pi$-periodic supercurrent is compatible with the presence of two gapless Andreev doublets. By changing the electron density, we find that the observed effect is predominant near the expected quantum spin Hall regime. In contrast, a non-topological HgTe quantum well is found to a conventional Shapiro response. Second, the response of the critical current to a perpendicular-to-plane magnetic field provides information on the spatial dependence of the current density (Fig.\ref{Fig:1TheoryDevice}D). When a junction is dominated by planar bulk modes, the uniform flow through the plane of the quantum well results in a standard Fraunhofer pattern (illustrated by a blue line in Fig.\ref{Fig:1TheoryDevice}D). When current flows only on the edges, a dc SQUID response is expected\cite{Hart2014} (red line). Our measurements exhibit a crossover between these two regimes suggesting that the $4\pi$-periodic current indeed flows along the edges (Fig.\ref{Fig:4Fraunhofer}). Additionally, in the SQUID regime, strong modulations of the odd lobes are observed, yielding an apparent doubling of the periodicity in the magnetic flux from $\Phi_0$ to $2\Phi_0$. Together, these sets of features strongly point towards the existence of topological gapless Andreev bound states with $4\pi$-periodicity flowing on the edges of the sample. Fig.\ref{Fig:2ExpDevice}C summarizes the evolution of these signatures on the gate voltage axis, and we detail our experimental observations in the remainder of the article.

The junctions are fabricated from epitaxially grown quantum wells of HgTe of thickness $d=\SI{8}{\nano\meter}$, sandwiched between barrier layers of Hg$_{0.3}$Cd$_{0.7}$Te on a CdZnTe substrate. In such wells (with a thickness larger than a critical thickness $d>d_c=\SI{6.3}{\nano\meter}$ \cite{Bernevig2006}), the existence of topological edge channels in the absence of magnetic field has been predicted and experimentally demonstrated via resistance quantization \cite{Konig2007}, non-local transport \cite{Roth2009}, spin polarization measurements \cite{Bruene2012} and scanning-SQUID imaging\cite{Nowack2013}. The layout of the Josephson junctions is presented in Fig.\ref{Fig:2ExpDevice}A. A rectangular mesa of HgTe sandwiched in Hg$_{0.3}$Cd$_{0.7}$Te is first defined. After locally etching the cap layer, aluminium contacts are then deposited in-situ on the HgTe layer, using standard evaporation and lift-off techniques. A metallic gate (Au) is placed between the Al contacts to control the electron density. The Al superconducting stripes have a width of \SI{1}{\micro\meter}. The HgTe mesa has a width of \SI{4}{\micro\meter}, corresponding to the width of the weak link. With this design, the overlap of edge channels on opposite edges is suppressed as the estimated edge channel\cite{Zhou2008,Bruene2010} width is around \SI{200}{\nano\meter}. The length of the junction presented in this paper is nominally $L=\SI{400}{\nano\meter}$. Given an estimated mean free path $l>\SI{2}{\micro\meter}$, and a coherence length $\xi\simeq\SI{600}{\nano\meter}$, we estimate that our junction is in an intermediate length regime $L\sim\xi< l$ (see Supplementary Information).

A typical $I$-$V$ curve (measured at 30 mK) is presented in Fig.\ref{Fig:2ExpDevice}B. The junction exhibits a Josephson supercurrent with a critical current $I_c=\SI{1.1}{\micro\ampere}$ (here for a gate voltage $V_g=\SI{0}{\volt}$). Hysteresis is observed between forward and reverse sweeps, with a retrapping current $I_r<I_c$. For voltages larger than the energy gap of the aluminium, the $I$-$V$ curve reaches an asymptote that does not go through the origin (grey line). The slope indicates the normal state resistance of the device $R_n$, while the intercept is the excess current $I_{exc}$. The excess current\cite{Blonder1982} stems from Andreev reflections in an energy window near the superconducting gap. It thus signals the presence of Andreev reflections at the S-TI interfaces and underlines the quality of our junctions. In order to identify the QSH regime, it is instructive to plot the normal state resistance $R_n$ and the critical current $I_c$ as a function of the gate voltage $V_g$ (Fig.\ref{Fig:2ExpDevice}C). We observe three regimes. For gate voltages between $V_g=\SI{-1.1}{\volt}$ and \SI{0}{\volt}, $R_n$ is low (below \SI{300}{\ohm}) and $I_c$ is large (above \SI{200}{\nano\ampere}) thus characterizing high mobility $n$-type conduction. For gate voltages below $V_g=\SI{-1.7}{\volt}$, the normal state resistance decreases again slowly, indicating the $p$-conducting regime. Due to a lower mobility in this region, the critical current $I_c$ lies below \SI{50}{\nano\ampere}. In between, $R_n$ exhibits a peak with a maximum around $\SI{1.5}{\kilo\ohm}$ (for $V_g=\SI{-1.45}{\volt}$), for which $I_c$ is almost suppressed.  This indicates the region where the QSH edge states should be most visible. However, the peak value of $R_n$ is lower than the expected quantized value $h/2e^2\simeq\SI{12.9}{\kilo\ohm}$, thus suggesting the presence of residual bulk modes in the junction\cite{Hart2014}, possibly due to Al diffusion\cite{Capper1994} into the HgTe material. Additionally, local $n$-doping caused by Al may result in $p$-$n$ barriers at the interface between the Al capped- and gated areas. This would contribute to the two-point normal state resistance $R_n$ and could obscure a correct identification of the QSH transition.

The presence of gapless Andreev bound states can in principle be detected via the $4\pi$-periodic contribution to the supercurrent. In practice such detection in dc transport can be complicated by additional contributions from conventional $2\pi$-periodic modes\cite{Finck2014,Galletti2014} carrying a current $I_{2\pi}\sin\varphi$ and by relaxation processes which can restore a $2\pi$-periodic supercurrent\cite{Kwon2004, Pikulin2012, Badiane2013}. In order to reveal possible the $4\pi$-periodic characteristics, the dynamics of the junction is most conveniently probed by studying the ac-Josephson effect. In a previous work\cite{Wiedenmann2016}, we identified anomalous features in the Shapiro response of a weak link made of the 3D TI strained HgTe (namely a missing $1^{\rm st}$ Shapiro step) which we attributed to a fractional ac Josephson effect consistent with the presence of a $4\pi$-periodic Andreev doublet. In the present report, we apply the same method to the more interesting case of a 2D TI. The current-voltage characteristic of the sample is measured using a dc current bias with the addition of an rf driving current, coupled to the device via a nearby antenna\cite{Wiedenmann2016}. Fig.\ref{Fig:3Shapiro}A shows a series of histograms of the voltage distribution obtained when lowering the excitation frequency (at $V_g=\SI{-1.1}{\volt}$, see Supplementary Information for typical $I$-$V$ curves). For a high frequency excitation ($f=\SI{6.6}{\giga\hertz}$), we observe a 'conventional' sequence of Shapiro steps at $V=nhf/2e,n\in\mathbb{Z}$. As the frequency is lowered to $f=\SI{0.8}{\giga\hertz}$, we observe the progressive vanishing of all odd steps up to $n=9$. To our knowledge, the only mechanism resulting in the suppression of odd steps is the presence of a $4\pi$-periodic contribution in the supercurrent. While a pure $4\pi$-periodic supercurrent should lead directly to an even sequence of Shapiro steps (as a direct effect of the substitution $\varphi\to\varphi/2$ in the Josephson equations), it is in our experiment only visible at low rf frequency. Our junction may in practice contain both gapless bound states and a number of residual conventional modes such that the supercurrent $I_s$ could be written as $I_s(\varphi)=I_{4\pi}\sin\varphi/2 + I_{2\pi}\sin\varphi\,(+\rm higher\, harmonics)$. Even in the presence of a strong $2\pi$-periodic contribution, a $4\pi$-periodic response can be observed when the time dependence of the voltage $V$ to the current bias $I$ is most anharmonic, namely at a low frequency\cite{Dominguez2012,Wiedenmann2016} (see SI). Odd steps are then missing if the rf excitation frequency $f$ is lower than a frequency $f_{4\pi}=\frac{2eR_n I_{4\pi}}{h}$. An estimate of the crossover frequency $f_{4\pi}$ then yields the amplitude of the $4\pi$-periodic supercurrent $I_{4\pi}$. Estimating the crossover frequency $f_{4\pi}$ by a fully suppressed $n=3$ step, we find that $I_{4\pi}\simeq\SI{20}{\nano\ampere}$. The presence of two gapless Andreev bound states carrying a current of $\SI{10}{\nano\ampere}$ is compatible with the maximum current $\frac{e\Delta_{\rm i}}{\hbar}$ for the estimated induced gap $\Delta_{\rm i}\geq\SI{80}{\micro\electronvolt}$ (see SI) and the gap of the Al contacts ($\Delta_{\rm Al}\simeq \SI{150}{\micro\electronvolt}$). Apart from gapless Andreev states, gapped Andreev bound states with high-transparency could also result in a $4\pi$-periodic contribution in the supercurrent in the presence of Landau-Zener transitions at the avoided crossing $\varphi=\pi,3\pi,...$ \cite{Sau2012,Dominguez2012}. However Landau-Zener transitions have increasing probabilities with dc voltage or frequency. Given our observations of missing steps only at low frequencies, this possibility appears unlikely. Recently it has been proposed that Coulomb interaction and coupling of Andreev bound states to the continuum can result in a quantum spin Hall system in the appearance of $8\pi$-periodic Josephson effect \cite{Zhang2014}. In this experimental work, we observe no sign of such an effect in the response of our devices.

We now address the dependence of the Shapiro steps on the rf power. Fig.\ref{Fig:3Shapiro}B shows as an example the behaviour at $f=\SI{1.8}{\giga\hertz}$ (see Supplementary Information for additional data sets). A two-dimensional map of the voltage histogram is shown as a function of voltage and excitation power $P_{rf}$. For low power levels, the steps progressively appear (starting from low step indices) as the power is increased. While maxima are seen for $n=0,2,4$, the first and third Shapiro steps are fully suppressed (as illustrated previously by the histogram of Fig.\ref{Fig:3Shapiro}A taken along the red dashed line). As visible here, hysteresis is occasionally found to induce a weak asymmetry at low power, affecting the $n=\pm1$ step at low frequency and low power. For high power, an oscillatory pattern (reminiscent of Bessel functions in the voltage bias case\cite{Russer1972}) is observed.
However the pattern is deeply modified (when compared with that seen for higher frequency excitation) with dark fringes\cite{Wiedenmann2016} (white dotted lines) which develop from the suppression of the first and third maxima of the oscillations. In our opinion these features signal the progressive transformation from a $2\pi$- to a $4\pi$-periodic pattern with halved period of oscillations corresponding to a halved number of steps.

\begin{figure}[h!]
\centerline{\includegraphics[width=\textwidth]{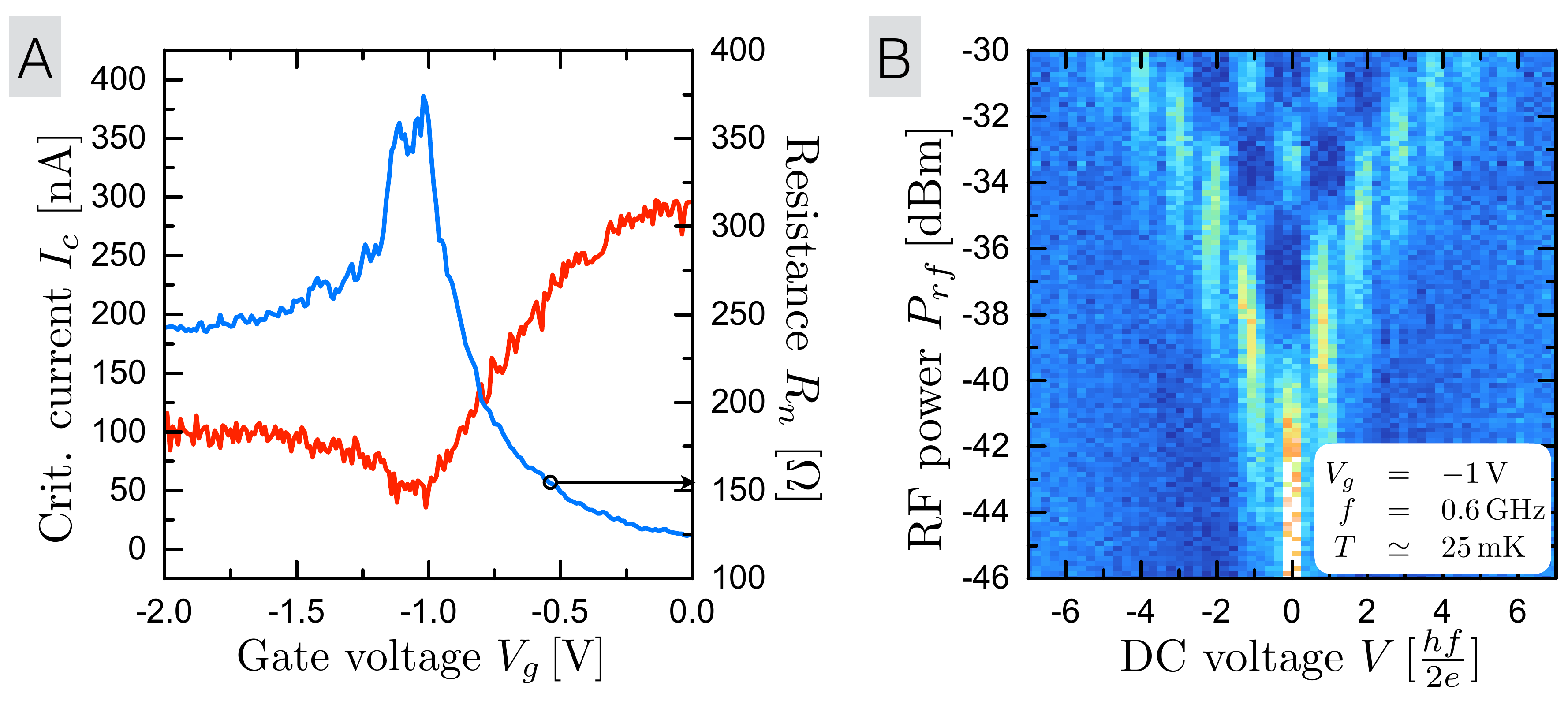}}
\caption{{\bf Shapiro response of a junction on a trivial quantum well} A) Critical current $I_c$ and normal state resistance $R_n$ of a non-topological device as a function of gate voltage $V_g$. B) 2D maps of the voltage distribution as a function of the voltage (in normalized units) and the RF excitation power. It is taken at gate voltage $V_g\simeq\SI{-1}{\volt}$ and frequency $f=\SI{0.6}{\giga\hertz}$. At this frequency close to our resolution limit, all steps are still visible.} \label{Fig:TrivialShapiro}
\end{figure}

\begin{figure}[h!]
\centerline{\includegraphics[width=1\textwidth]{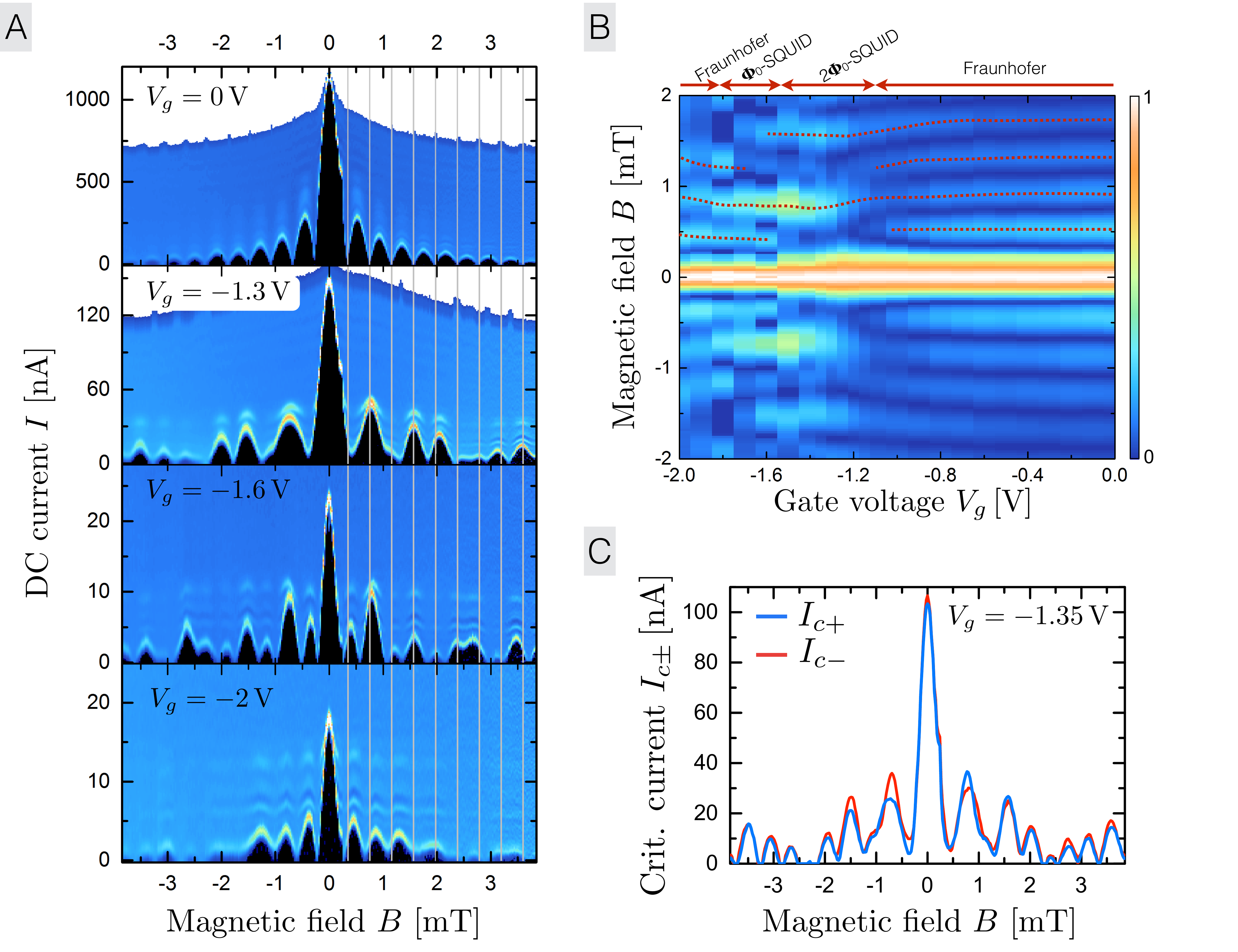}}
\caption{{\bf Response of the critical current to a magnetic field -} A) Maps of differential resistance $dV/dI$ as function of magnetic field $B$ and dc current bias $I$ for four different gate voltages. For $V_g=\SI{0}{\volt}$ and \SI{-2}{\volt}, the patterns are close to Fraunhofer diffraction ones, while at $V_g=\SI{-1.3}{\volt}$ and $V_g=\SI{-1.6}{\volt}$, a SQUID-like pattern is observed. Vertical grey lines indicate minima of the Fraunhofer pattern where $\Phi$ is a multiple of $\Phi_0$, that correspond to maxima of the SQUID pattern. For $V_g=\SI{-1.3}{\volt}$, the first and third side lobes are suppressed, yielding an apparent doubling of the period of the interference. B) Map of normalized critical currents as function of magnetic field $B$ and gate voltage $V_g$. The disappearance of first and third lobes as well as shifts in position of second and fourth lobes are highlighted by red dashed lines. C) Critical current as a function of magnetic field at $V_g=\SI{-1.35}{\volt}$, for the two sweep directions (positive as a blue line, negative as a red line), following the symmetry relation $I_{c+}(B)=I_{c-}(-B)$.} \label{Fig:4Fraunhofer}
\end{figure}

Next we investigate the dependence of the Shapiro response on gate voltage $V_g$. A general observation is that odd steps vanish at low frequencies for a very wide range of gate voltages, from about $V_g\geq\SI{-1.3}{V}$ up to \SI{+0.5}{V}. The visibility of an even sequence is in agreement with the previously introduced criterion $f<f_{4\pi}$, and demonstrates that a $4\pi$-periodic contribution is present in addition to a conventional $2\pi$-periodic component. The latter most likely originates from the bulk signaling the expected coexistence of topological edge states with modes from the conduction band \cite{Dai2008}. The $4\pi$-periodic modes are progressively unveiled as the number of bulk modes is decreased for negative voltages. In Fig.\ref{Fig:3Shapiro}C, we show a plot of the ratios $Q_{12}$ (resp. $Q_{34}$) of the maximum amplitude of the $n=1$ to $n=2$ steps ($n=3$ to $n=4$ steps respectively) for measurements at $f=\SI{3.1}{\giga\hertz}$. Simulations using the RSJ model \cite{Dominguez2012,Wiedenmann2016} predict ratios $Q_{i,i+1}$ which are close to unity for a conventional junction, as indicated by the shaded grey area. Ratios approaching zero indicate the suppression and disappearance of the odd Shapiro step. Both ratios $Q_{12}$ and $Q_{34}$ indicate that the visibility of the even sequence of steps is improved between $V_g=\SI{-1.3}{V}$ to \SI{-0.9}{\volt}. Thus, the ac response of our junctions strongly signals the presence of a strong $4\pi$-periodic contribution from the supercurrent appearing more clearly in this range close to the QSH transition. Furthermore, we emphasize that these observations are fully consistent with our previous work on the 3D topological insulator strained HgTe\cite{Wiedenmann2016}, in which a single gapless Andreev doublet is expected \cite{Tkachov2013}. However observation is in this case less favorable due to the presence of a greater number of conventional modes, displacing the crossover to lower frequencies in which the visibility of Shapiro steps is reduced, making it difficult to see more than one missing odd step.
To assess the topological origin of the $4\pi$-periodic supercurrent, we now briefly examine a narrower HgTe quantum well (thickness of $d=\SI{5}{\nano\meter}<d_c$, see SI), that does not exhibit the quantum spin Hall effect\cite{Konig2007}. The measurement of the critical current $I_c$ enables the identification of the $n$- and $p$- conduction regimes though the gap in between around $V_g\simeq\SI{-1}{\volt}$ is not very pronounced (see Fig.\ref{Fig:TrivialShapiro}A). When measuring the Shapiro response to an rf excitation, we do not observe any missing odd step for any of the gate voltage, neither in $n$- nor $p$- regimes, nor close to the gap. As an example, we show in Fig.\ref{Fig:TrivialShapiro}B a measurement taken close to the gap at $V_g=\SI{-1}{\volt}$ and $f=\SI{0.6}{\giga\hertz}$. For such a frequency close to our detection limit, all steps are still visible. 

Finally we detail the response of the junction to a magnetic field perpendicular to the plane of the junction, and show that the previous observations are compatible with edge transport in the same range. With application of a magnetic field, the superconducting phase difference $\varphi$ becomes position dependent\cite{Tinkham2004}. This in turn reveals properties of the spatial supercurrent distribution through modulations of the critical current $I_c$ with a period given by the magnetic flux quantum $\Phi_0=h/2e$. In Fig.\ref{Fig:4Fraunhofer}A, we present the differential resistance $dV/dI$ as a function of dc current $I$ and magnetic field $B$ for different gate voltages corresponding to four different behaviours that we identify. First for $V_g=\SI{0}{\volt}$, the junction exhibits a conventional Fraunhofer pattern of the critical current versus magnetic field, that rapidly decays as the magnetic field increases. In this regime, the electron density is high and current flows uniformly in the 2D plane of the quantum well. We obtain a period of circa \SI{0.41}{\milli\tesla} which, given the dimensions of our device, corresponds to an effective area $W(L+2\lambda)=\SI{5.1}{\square\micro\meter}$ yielding a penetration length $\lambda\simeq\SI{430}{\nano\meter}$ (similar to Ref.\onlinecite{Hart2014}). The minima are then identified as points where the magnetic flux $\Phi$ through the junction is a multiple of the flux quantum $\Phi_0$ ($\Phi=n\Phi_0,n\in\mathbb{Z}$). As the gate voltage is decreased, the critical current decreases, and the diffraction pattern is similar to that of a (dc) SQUID for $V_g=\SI{-1.3}{\volt}$ and $V_g=\SI{-1.6}{\volt}$. The presence of maxima at multiples $\Phi=n\Phi_0,n\in\mathbb{Z}$ is a signature of a SQUID-like behavior and demonstrates that a sizeable part of the supercurrent flows along the edges of the sample\cite{Hart2014}, as expected in the presence of QSH edge channels. In addition, the narrowing of the central lobe is clearly visible at $V_g=\SI{-1.6}{\volt}$. However, very strong odd/even modulations are observed  in both cases : the first and third lobes are substantially smaller than the second and fourth. In particular at $V_g=\SI{-1.35}{\volt}$, the first and third lobes are completely suppressed, yielding an apparent doubling of the period (from $\Phi_0$ to $2\Phi_0$) at low fields, before a conventional period is recovered for larger fields. Finally for more negative gate voltages ($V_g=\SI{-2}{\volt}$), the pattern progressively returns to a Fraunhofer one, with some strong distortions, especially at high fields. This suggests that the current flow returns to a two-dimensional configuration, with inhomogeneities likely to stem from the lower mobility of the charge carriers in this gate voltage range. 
In Fig.\ref{Fig:4Fraunhofer}B, we present a two-dimensional plot of the normalized critical current $I_c(B)/I_c(B=0)$ when the gate voltage $V_g$ and magnetic field $B$ are varied. For $V_g=\SI{0}{\volt}$ to $V_g=\SI{-0.8}{\volt}$ the pattern remains close to a Fraunhofer pattern, but the first and third lobes progressively disappear and are missing between $\SI{-1}{\volt}$ and $\SI{-1.5}{\volt}$, as emphasized by the thin red guides to the eye. A progressive shift of the position of the second and fourth maxima is also visible towards $\Phi/\Phi_0\simeq2$ and $4$ in the SQUID-like regime. For $\SI{-1.6}{\volt}$, the first and third lobes reappear. When driving the gate from  $V_g=\SI{-1.6}{\volt}$ to $V_g=\SI{-2}{\volt}$, a standard Fraunhofer pattern is progressively recovered.

It is tempting to associate our observations of the anomalous doubled period to a SQUID-like pattern $|\cos(\frac{\pi\Phi}{2\Phi_0})|$ with periodicity $2\Phi_0$ originating from the $4\pi$-periodicity previously identified. But dc measurements are sensitive to relaxation processes that in principle restore a conventional $\Phi_0$ periodicity\cite{Fu2008,Houzet2013,Lee2014}, and it is unlikely that our devices are free of quasiparticle poisoning during the timescale of the experiment. Several models\cite{Baxevanis2015,Tkachov2015} have explained odd-even patterns and small deviations in a previously observed SQUID pattern\cite{Pribiag2015} via skewed current phase relations or additional coupling between edges. Neither the lobe pattern nor the effect of temperature (see Supplementary Information) favour such models.  An alternative mechanism to explain the interference pattern is the interplay of Zeeman effect and spin-orbit coupling, which should occur in our material system\cite{Dolcini2015,Rasmussen2015}. The interference pattern shows a peculiar symmetry relation of the critical current :  $I_{c+}(B)= I_{c-}(-B)$, where $\pm$ indicates the sweep direction of the bias current and $\pm B$ the magnetic field direction (see Fig.\ref{Fig:4Fraunhofer}C). In contrast, it is asymmetric both in magnetic field and sweep direction. Further investigation in a more suitable geometry is required to clarify the role of these mechanisms.

Finally, we return to Fig.\ref{Fig:2ExpDevice}C. The response to rf irradiation strongly suggests the presence of a $4\pi$-periodic supercurrent in the device with a contribution compatible with two modes. It is most visible when the bulk bands are depleted (as indicated by the Fraunhofer interference pattern). In this region, the current flow is mostly along the edges of the sample (as indicated by the SQUID features, with possible indications of spin-orbit and Zeeman effects). Though the QSH regime is not clearly identified by its quantized conductance, it appears that the $4\pi$-periodic contribution is also detected in the whole $n$-conduction band but is rapidly suppressed when driving the gate voltage towards the $p$-conduction regime. This suggests that the $4\pi$-periodic edge modes coexist in parallel with bulk modes of the conduction band, consistently with previous predictions\cite{Dai2008} and observations in our material system\cite{Nowack2013,Hart2014}. In contrast, a Josephson junction made of a topological trivial quantum well exhibits a conventional Shapiro response. Together, this set of observations strongly point towards the existence of topological gapless Andreev bound states predicted by Fu and Kane\cite{Fu2009} in Josephson junctions produced on the well-characterized QSH insulator HgTe. Though further developments are required to fully comprehend the richness of the observed phenomena, Josephson junctions in HgTe quantum wells and at zero magnetic field appear promising for the future realization of Majorana end states and possibly scalable Majorana qu-bits.
\newpage
{\bf Acknowledgments:}

We thank V. Hock, L. Maier for technical assistance and gratefully acknowledge L.~Glazman, Y.~Peng, F.~von~Oppen, E.M. Hankiewicz, G. Tkachov and B. Trauzettel for enlightening discussions. This work is supported by German Research Foundation (Leibniz Program, DFG-Sonderforschungsbereich 1170 'Tocotronics' and DFG-Schwerpunktprogramme 1666), the Elitenetzwerk Bayern program “Topologische Isolatoren”. R.S.D. acknowledges support from Grants-in-Aid for Young Scientists B (No. 26790008). T.M.K. is financially supported by the European Research Council Advanced grant No.339306 (METIQUM). E.B., T.M.K. and L.W.M. gratefully thank the Alexander von Humboldt foundation for its support.

\renewcommand{\thefigure}{S\arabic{figure}}
\setcounter{figure}{0}

\clearpage
\begin{center}

\Large{Gapless Andreev bound states in the quantum spin Hall insulator HgTe\\
--\\
Supplementary Material}
\normalsize
\end{center}

{\bf In this supplementary information, we present further measurements and detail several aspects not discussed in the main text. In particular, we provide raw measurements of the $I$-$V$ curves in the presence of rf irradiation, that explicitly show the absence of multiple odd steps. Finally the temperature dependence of the SQUID pattern with doubled periodicity is also described.}

\section{Response to an rf excitation}

\subsection{Characterization measurements and typical length scales}

The mobility and charge density are evaluated from a Hall-bar produced separately from the same wafer as the junctions. The extracted density and mobility are typically $n_e= \SI{2.2e11}{\per\centi\meter\squared}$ and $\mu=\SI{3e5}{\centi\meter\squared\per\volt\per\second}$. From these values, we calculate a mean free path of $l\simeq \SI{2.4}{\micro\meter}$. Assuming an induced gap $\Delta_{\rm i}\lesssim\Delta_{\rm Al}$, the natural coherence length $\xi=\frac{\hbar v_{\rm F}}{\pi\Delta_{\rm i}}$ is typically around \SI{600}{\nano\meter} in our system. The junction is expected to be in the ballistic limit with $L\sim\xi<l$.
\begin{figure}[h!]
\centerline{\includegraphics[width=1\textwidth]{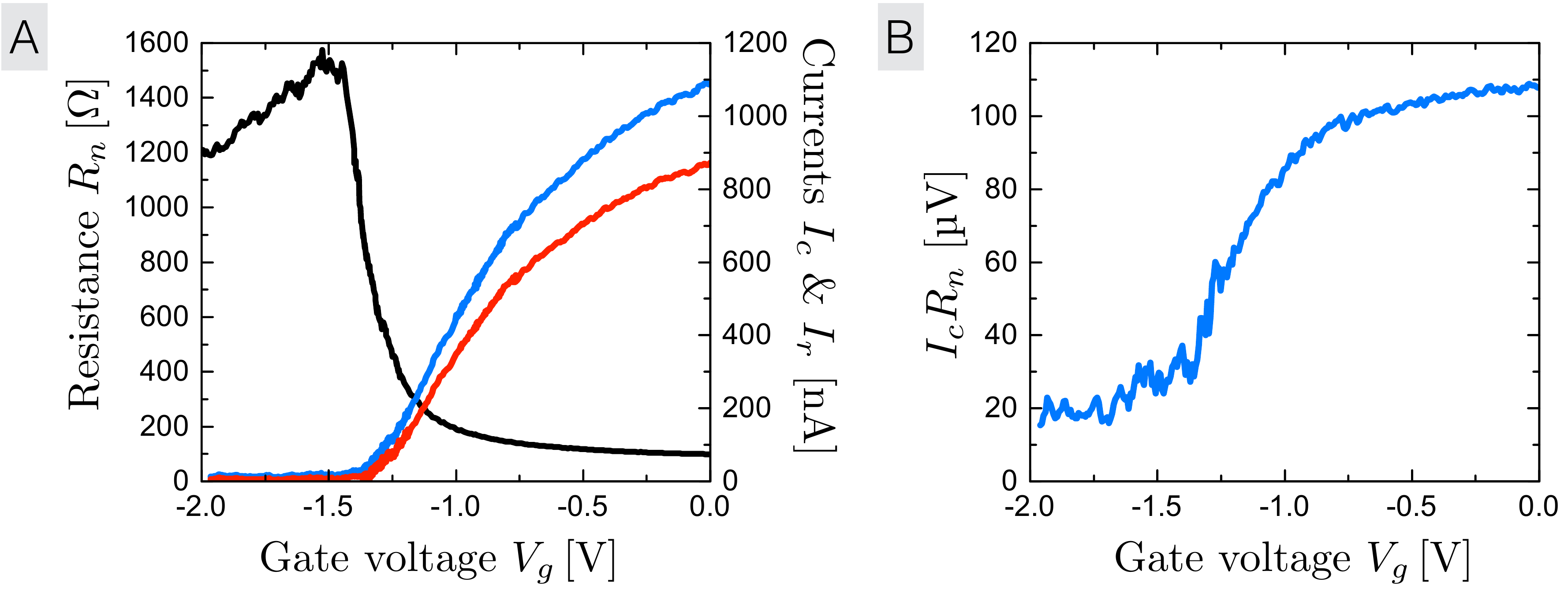}}
\caption{{\bf Parameters of the junction} A) Normal state resistance $R_n$, critical and retrapping current $I_c$ and $I_r$, as a function of gate voltage $V_g$. The difference between $I_c$ and $I_r$ reflects the weak hysteresis in the $I$-$V$ curve shown in the main text. B) $I_cR_n$ product as a function of gate voltage $V_g$. It approaches the Al gap (\SI{170}{\micro\volt}) for a gate voltage between \SI{-1}{\volt} and \SI{0}{\volt}, but decreases when $V_g<\SI{-1}{\volt}$.} \label{Fig:IcRnProduct}
\end{figure}
For the sake of completeness, we provide in Fig.\ref{Fig:IcRnProduct} the parameters measured on the junction presented in the main text. The critical current $I_c$ and normal state resistance $R_n$ (extracted for voltages larger than the Al gap) have already been presented in the main text. Additionally we show here the retrapping current $I_r$ and the $I_cR_n$ product. The latter saturates around \SI{110}{\micro\volt} for $V_g$ between \SI{1}{\volt} and \SI{0}{\volt}, but decays when the gate voltage is driven below \SI{-1}{\volt}. It yields a lower bound on the induced gap $I_cR_n\leq \Delta_{\rm i}/e$ compatible with the Al superconducting gap (around $\Delta_{\rm Al}\simeq\SI{170}{\micro\volt}$ for Al).

The junction realized in a narrower quantum well exhibits slightly lower mobility $\mu=\SI{1.5e5}{\centi\meter\squared\per\volt\per\second}$. Though the gap feature is not strongly visible in the presented Josephson junction (probably originating from local diffusion and doping due to the Al electrodes), one can verify that the characterization Hall-bar exhibits a clear insulating behavior in the gap.

\subsection{Examples of $I$-$V$ curves in the presence of rf irradiation}

\begin{figure}[h!]
\centerline{\includegraphics[width=0.55\textwidth]{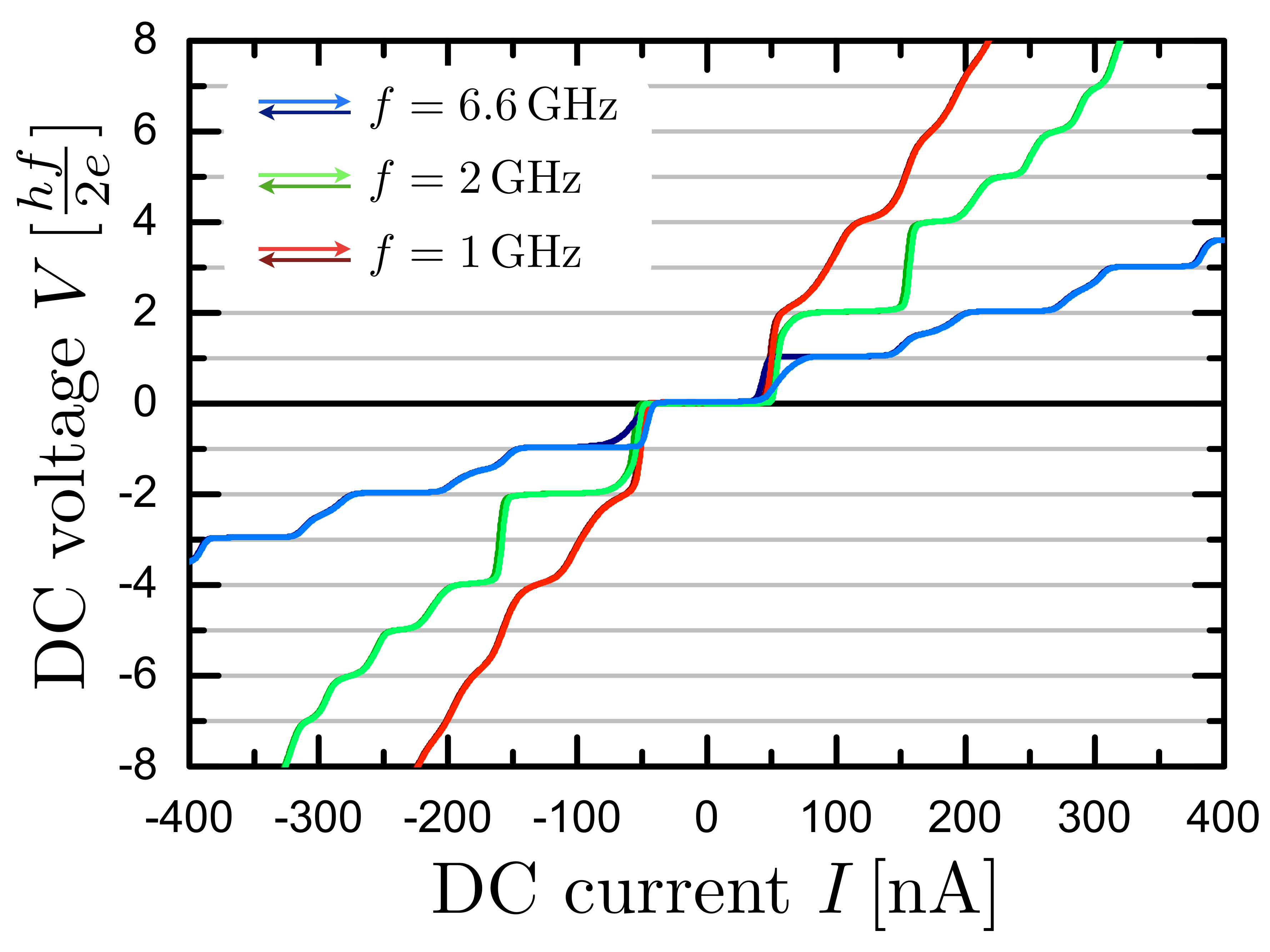}}
\caption{{\bf Typical $I$-$V$ curves in the presence of rf excitation} For three different excitation frequencies and for a gate voltage $V_g=\SI{-1.1}{\volt}$, the $I$-$V$ curves exhibit a different sequence of Shapiro steps. For $f=\SI{6.6}{\giga\hertz}$, all integer Shapiro steps are visible ($n=0,\pm1,\pm2,...$). Subharmonic steps at $n=3/2, 5/2,..$ are also weakly present. A weak asymmetry is observed between forward and reverse sweep directions, signaling a hysteresis in this region. For $f=\SI{2}{\giga\hertz}$, the first and third step are absent, but all steps are visible beyond $n=4$. In this case, the hysteresis is not visible anymore. For $f=\SI{1}{\giga\hertz}$, odd steps $n=\pm1,\pm3,\pm5$ are missing. At the same time, the steps (whose amplitude $hf/2e$ gets smaller) become progressively smoothed, making measurements for frequencies $f\leq\SI{0.8}{\giga\hertz}$ difficult.} \label{Fig:ShapiroIVcurves}
\end{figure}

In this section, we present typical $I$-$V$ curves measured in the presence of rf excitation (Fig.\ref{Fig:ShapiroIVcurves}). Multiple Shapiro steps are visible (up to $n\geq 10$), and reach the expected discrete values $V_n=nhf/2e, n\in\mathbb{Z}$ with a good accuracy. In the presented curves, all integer steps are visible at $f=\SI{6.6}{\giga\hertz}$. Additionally, subharmonic steps are weakly seen at half-integer values $n=3/2,5/2,...$. Such subharmonic steps are commonly reported and stem from the presence of capacitive effects or higher harmonics in the current phase relation\cite{Valizadeh2008}. We observe such subharmonic steps only at high frequencies, in a regime in which all integer steps are present. At lower frequencies, we observe the disappearance of the $n=\pm1,\pm3$ at $f=\SI{2}{\giga\hertz}$, and $n=\pm1,\pm3,\pm5$ at $f=\SI{1}{\giga\hertz}$. 

As the excitation frequency $f$ is lowered, the overall visibility of the steps is progressively reduced, as both the step height (proportional to $f$) and the step amplitude (on the current axis, see Ref.\onlinecite{Russer1972}) decrease with $f$. The $I$-$V$ curves measured for both sweep directions are presented in Fig.\ref{Fig:ShapiroIVcurves}. A weak asymmetry is observed on the first step, as a remnant of the hysteresis observed in the absence of rf excitation. It is seen to only affect the first step at low rf power.

\subsection{Frequency dependence}

\begin{figure}[h!]
\centerline{\includegraphics[width=0.7\textwidth]{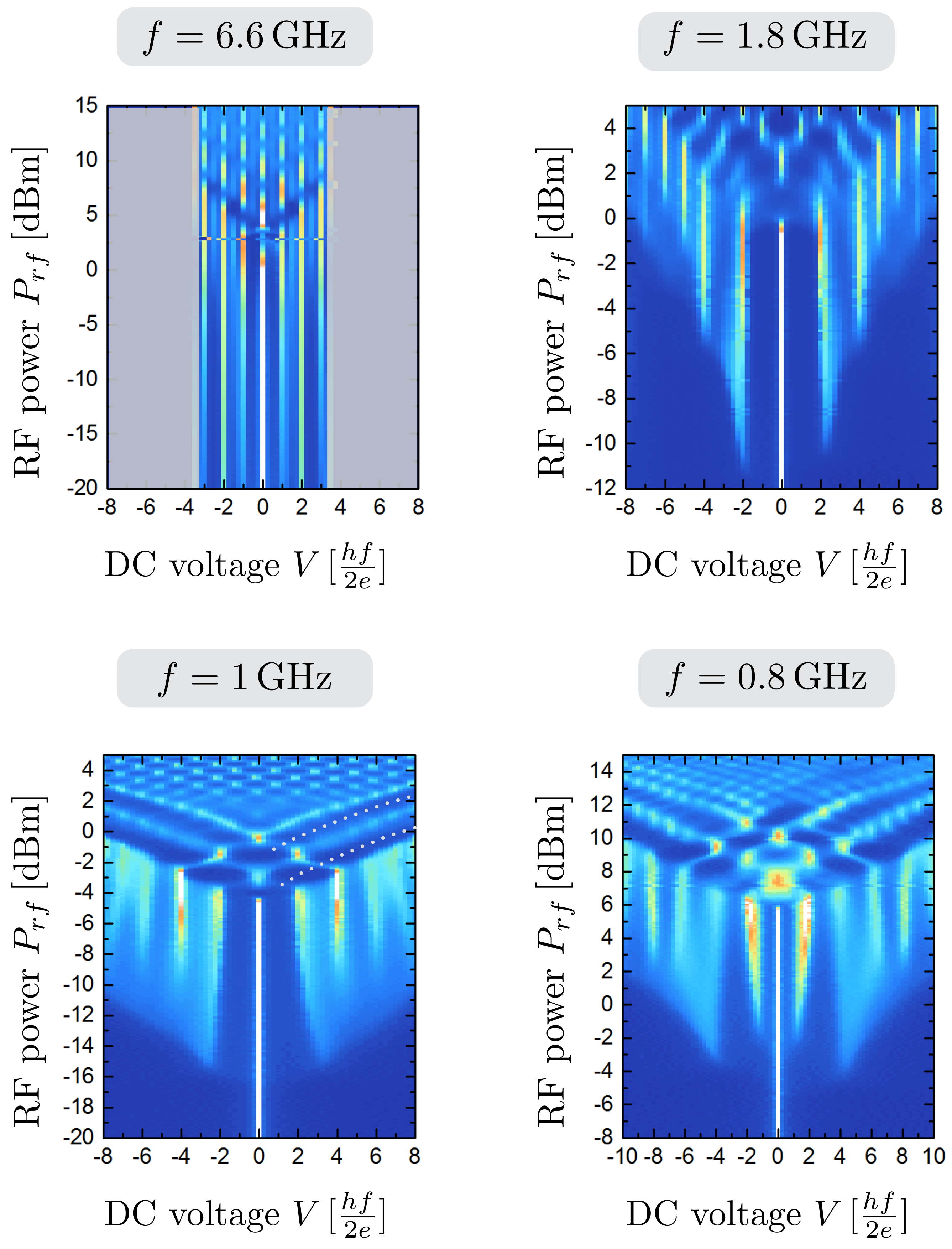}}
\caption{{\bf 2D maps of the voltage distribution} For four different excitation frequencies, the 2D maps reveal the presence of missing odd steps in the low power regime. In the high power regime, one observes the progressive suppression of odd lobes (first and third) as the frequency decreases, resulting in dark fringes (signaled by grey dotted lines).} \label{Fig:Shapiro2D}
\end{figure}

In Fig.\ref{Fig:Shapiro2D} we provide complete 2D plots of the voltage histograms for different frequencies presented in Fig.3A of the main article. As previously demonstrated, multiple odd steps vanish when the frequency is decreased. At $f=\SI{6.6}{\giga\hertz}$, all steps are visible, though one can already notice that the $n=1$ and 3 steps are weakened. At $f=\SI{1.8}{\giga\hertz}$, steps $n=1$ and 3 are fully suppressed, the $n=5$ vanishes as well at $f=\SI{1}{\giga\hertz}$, and finally the $n=7$ and 9 at $f=\SI{0.8}{\giga\hertz}$. Additionally, the high power oscillatory (reminiscent of Bessel functions\cite{Russer1972}) regime also evolves as $f$ decreases. While all oscillations are present at $f=\SI{6.6}{\giga\hertz}$, the first and third oscillations are suppressed at 1 and \SI{0.8}{\giga\hertz}, indicating that the oscillatory patterns also progressively turns into the one expected for a $4\pi$-periodic system, with half of the oscillations suppressed.
A few anomalies are visible in these maps. First a weak asymmetry is visible at low power for $f=\SI{1.8}{\giga\hertz}$. This is a consequence of the hysteresis previously mentioned, suppressed by a weak rf power and only seen to affect the first step. Second, the visibility of some of the Shapiro steps is sometimes poor (for example the $n=2$ step at $f=\SI{0.8}{\giga\hertz}$. The origin is not clear but we speculate on the possibility of complex and partially instable dynamics due to the presence of both $2\pi$- and $4\pi$-periodic components in the supercurrent.

\subsection{Summary of simulations with the Resistively Shunted Junction (RSJ) model}

In this section we briefly recall the results of numerical simulations carried out in the framework of the RSJ model. For a more extensive presentation, we refer the reader to the Supplementary Information of Reference \onlinecite{Wiedenmann2016}. In this framework, the junction is modeled by its current phase relation (CPR) $I_S(\varphi)$ together with a resistive shunt that represents the environment. When combined with the Josephson equation, one then readily obtains a first order differential equation on the superconducting phase difference $\varphi(t)$ that can be solved and consequently yields the voltage $V(t)=\frac{\hbar}{2e}\frac{d\varphi}{dt}$.

Focusing on the radiation-induced Shapiro steps, we believe that the essential features of the dynamics are properly captured in this RSJ model. For a more detailed analysis of the voltage- (or energy-dependence) of the I-V curves, microscopic details are crucial. We have simulated the effects of several CPR (simple $I_{2\pi}\sin\varphi + I_{4\pi}\sin\varphi/2$, or sum over modes with different transmissions) and our conclusions are the following.
\begin{itemize}
\item A $4\pi$-periodic ($\sin\varphi/2$) contribution is in all cases necessary to get missing steps.
\item All odd steps progressively disappear as the frequency $f$ is decreased, and the crossover frequency can be identified as predicted\cite{Dominguez2012} as $f_{4\pi}=\frac{2eR_nI_{4\pi}}{h}$ .
\item The exact form of the gapped $2\pi$-periodic modes in the CPR matters extremely marginally on the observed disappearance of the odd steps. Only the presence of a $4\pi$-periodic contribution is required to observe the disappearance of odd steps.
\end{itemize}
The RSJ model provides a minimal tool to simulate the dynamics of our system, using microscopic inputs from more elaborate theories, in the absence of a full microscopic model of the Josephson dynamics.

\section{Response to a magnetic field}
\begin{figure}[h!]
\centerline{\includegraphics[width=1\textwidth]{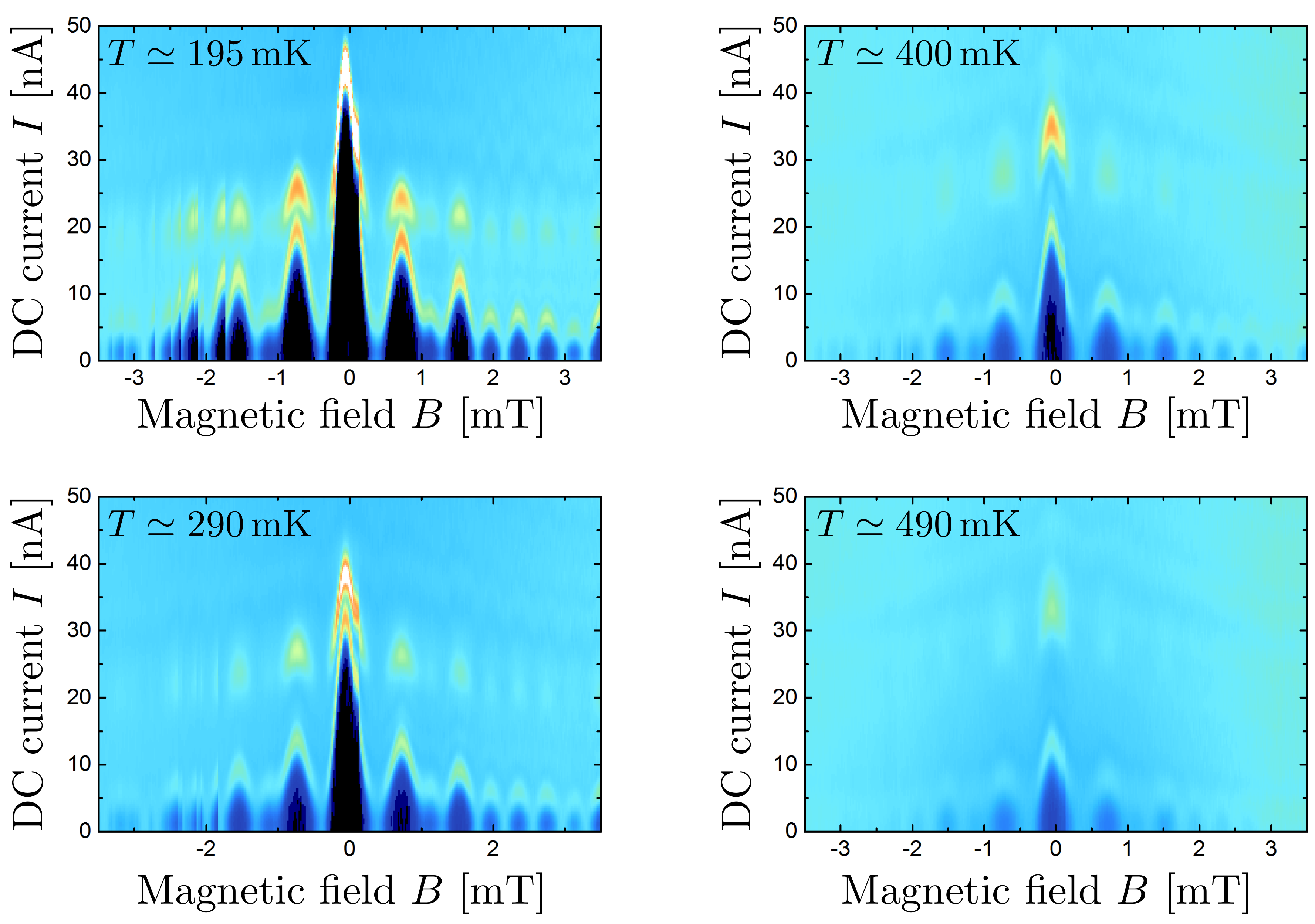}}
\caption{{\bf Temperature dependence of the response to magnetic field -} For four different temperatures, maps of differential resistance $dV/dI$ are presented as function of magnetic field $B$ and DC current bias $I$. The measurement is taken at $V_g=\SI{-1.35}{\volt}$, for which a SQUID pattern with strong odd/even modulations is observed. While the temperature alters the overall amplitude of the critical current, the interference pattern remains mostly unchanged.} \label{Fig:FraunhoferTdep}
\end{figure}
In Fig.\ref{Fig:FraunhoferTdep}, we show maps of the differential resistance $dV/dI$ as a function of the perpendicular magnetic field $B$ and DC current bias $I$. Measurements are taken at a gate voltage $V_g=\SI{-1.35}{\volt}$, for which the odd/even modulations in the interference pattern are maximal, thus giving an apparent periodicity of $2\Phi_0$. At $T\simeq\SI{195}{\milli\kelvin}$, the first and third side lobes are almost completely suppressed\footnote{For negative magnetic fields, jumps in the magnetic fields are visible, probably originating from trapped fluxes in the superconducting coil, but the positive side is unaltered.}. As the temperature is increased, we observe a progressive reduction of the amplitude of the measured critical current, but the interference pattern retains a very strong asymmetry between odd and even side lobes. This observation seems to exclude the mechanisms proposed by Tkachov {\it et al.} and Baxevanis {\it et al.} (Ref.  \onlinecite{Tkachov2015,Baxevanis2015}) to explain such asymmetries, as both models predict the disappearance of this asymmetry with increasing temperatures.

\bibliographystyle{unsrt}

\bibliography{BibShapiro2D.bib}

\end{document}